\documentstyle[12pt]{article}
\begin{document}
\baselineskip=24pt
\begin{center}
{\Large {\bf A fresh look at the Bohr - Rosenfeld analysis and a proof of a conjecture of Heisenberg}}\\
~\\
~\\
\baselineskip=12pt
{\bf H. Gopalkrishna Gadiyar}\\
AU-KBC Centre for Internet \& Telecom Technologies\\
M.I.T. Campus of Anna University, Chromepet, Chennai 600 044, INDIA\\
E-mail: gadiyar@au-kbc.org 
\end{center}

\vspace{3cm}
\begin{center}
{\bf Abstract}
\end{center}

\baselineskip=12pt

Bohr and Rosenfeld carried out an analysis of the consequences of field theory commutation relations. In this note the analysis is sharpened. A conjecture of Heisenberg that volume is quantized is shown to be a consequence of the second quantization of gauge fields. A way to generalize the equations of physics to include the Planck length is indicated. 

\newpage

In this paper we investigate the relationships between Bohr - Sommerfeld quantization, Bohr - Rosenfeld analysis of field theory commutators and the Heisenberg commutation relations. The analysis does not use the language of homology, cohomology, and cocycles ([3], [4], [5]) though the connection to abstract mathematics would be obvious to experts. This is done so that the working physicist appreciates the argument. 

We essentially study the relationship between the Heisenberg commutation relations and Bohr-Sommerfeld integrality condition. The key idea is that if there is a Bohr-Sommerfeld type condition one should apply the correspondence principle and replace the condition by a Heisenberg type commutator. In the next few paragraphs we will collect the Heisenberg commutation relations of quantum mechanics, quantum electrodynamics and quantum gravity. We reinterpret the loop approach [12] to quantum gravity as merely going from the Heisenberg algebra to the Weyl-Heisenberg group. At the group level one recovers the Bohr-Sommerfeld old quantum conditions. In the case of quantum mechanics nothing new is got. In the case of gauge fields the condition turns out to be the integrality of the Gauss linking number([7], [8]). In exact analogy the argument converts this ``old quantum condition" into a Heisenberg algebra where a commutation relation is proposed for length and area operators.

We quickly summarize the content of quantum physics required for our analysis ([9], [10], [11]). The equations obeyed by classical mechanics are
\begin{eqnarray*}
\dot{q}&=& \left \{ q, H \right \} \, ,\\
\dot{p}&=&\left \{ p, H \right \} \, ,
\end{eqnarray*}
where 
$$
\left \{ f,~g \right \} = 
\frac{\partial f}{\partial q} \frac{\partial g}{\partial q} - 
\frac{\partial f}{\partial p} \frac{\partial g}{\partial p} \, .
$$
Hence $\left \{ q,~p \right \}~=~1$.
In old quantum theory the Bohr-Sommerfeld condition  
$$
\frac{1}{2} \oint (p~dq ~-~q~dp)~=~n \, ,
$$
or equivalently, $\int \int dq~dp~=~n$ is an additional ad-hoc condition. In the full quantum mechanics of Heisenberg, the form of equations is preserved but $p$ and $q$ become operators with 
$$
[ \hat{q}, \hat{p} ] ~=~ i\hat{1} \, ,
$$
where 
$$ 
[ \hat{A}, \hat{B} ] ~=~ \hat{A} \hat{B}-\hat{B} \hat{A} \, .
$$ 
The standard notation of hatted objects for $q$-numbers and unhatted for $c$-numbers is followed.

For our purposes the commutation relations
\begin{eqnarray*}
~~ [ \hat{A}^j(x,t), \hat{A}_k(y, t) ]  &=& 0 \, , \\ 
~~ [ \hat{E}^j(x,t), \hat{E}_k(y, t)  ] &=& 0 \, , \\ 
~~ [ \hat{A}^j(x,t), \hat{E}_k(y, t)  ] &=& i \delta ^j_{k} \delta ^3(x-y \, , )
\end{eqnarray*}
with $j,~k~=~x,~y,~z$ are the analogue of quantum mechanical commutation relations for the electromagnetic field. 

Similar to the previous case with an additional complication of an extra symbol denoting a group index we have 
\begin{eqnarray*}
~~ [ \hat{A}^j_a(x,t), \hat{A}^b_k(y, t) ]  &=& 0 \, , \\ 
~~ [ \hat{E}^j_a(x,t), \hat{E}^b_k(y, t)  ] &=& 0 \, , \\ 
~~ [ \hat{A}^j_a(x,t), \hat{E}^b_k(y, t)  ] &=& i \delta ^j_{k} \delta ^b_a \delta ^3(x-y) \, , 
\end{eqnarray*}
with $j,~k~=~x,~y,~z$. $a,~b,~=~1,~2,~3$ are group indices in the case of the gravitational field.

Next we do what looks like a step backwards. Historically the Bohr-Sommerfeld quantum condition came first and Heisenberg derived his commutation relations later. Instead we construct the Weyl-Heisenberg group. This has the form
$$
U(p, q) = e^{i(q \hat{p} - p \hat{q})} \, .
$$
We construct the unitary operators which go from the origin to the point $(p,~q)$ then to $(p~+dp,~q~+~dq)$ and back. This would be an infinitesimal triangle which would be related to angular momentum. Mathematically
$$
U(p, ~q)~U(dp, ~dq)~U^{-1}(p~+~dp,~q+~dq)~=~e^{\frac{1}{2} (p~dq~-~q~dp)} \, .
$$
For a full closed path in phase space the Bohr-Sommerfeld relation will follow. That is, in mathematical terms the central extension at the level of the Heisenberg algebra leads to an integrality condition at the group level ([3], [4], [5]). 

To carry out the program we change two things in the Bohr-Rosenfeld analysis. Bohr and Rosenfeld worked with electric and magnetic fields and hence had derivatives of delta functions. We rectify this by using the vector potential and the electric field as can be seen from the earlier sections this makes the analogy with quantum mechanics very precise. Next we replace the Heisenberg algebra of the fields by the Weyl-Heisenberg group. This is done by taking the loop space version of electrodynamics. The objects are typically like
$e^{i\oint \vec{A}~d\vec{l}}$. We view $\vec{A}.d\vec{l}$ mathematically as a differential form and physically (in the spirit of Bohr and Rosenfeld) as a field $A_i$ coupling to a particle $dx_i$. This analysis is carried out by many people ([7], [8]) as it leads to a well known topological invariant namely the Gauss linking number. In other words
$$
\oint d\vec{l}(x) \int d\vec{S}(y) \delta ^3(x,y)= n \, .
$$
We now summarize the situation as follows: the second quantized field in 3+1 dimensions leads to a Bohr-Sommerfeld type ``old quantum theory" condition. This would mean that in analogy to the passage from Bohr-Sommerfeld rule to Heisenberg commutation relations there should be commutation relations
\begin{eqnarray*}
~~ [ \hat{l}^j(x), \hat{l}_k(y) ]  &=& 0 \, ,\\ 
~~ [ \hat{S}^j(x), \hat{S}_k(y)  ] &=& 0 \, ,\\ 
~~ [ \hat{l}^j(x), \hat{S}_k(y)  ] &=& L^3 \delta ^j_{k} \delta ^3(x-y)\, ,
\end{eqnarray*}
with $j,~k~=~x,~y,~z$. The commutation relations would lead to quantization of volume in analogy with the harmonic oscillator. Heisenberg had conjectured in [6] that volume should be quantized and this approach seems to lead to his conclusion. One could take advantage of the Hodge $^*$ operation in 3 dimensions and use the fact that $^*dx~=~dy~\wedge dz$ and cast the commutation relation in the form 
\begin{eqnarray*}
~~ [ \hat{x}^j(x), \hat{x}_k(x') ]  &=& 0 \, ,\\ 
~~ [ \hat{x}^{j*}(x), \hat{x}^*_k(x')  ] &=& 0 \, ,\\ 
~~ [ \hat{x}^j(x), \hat{x}^*_k(x')  ] &=& L^3 \delta ^j_{k} \delta ^3(x-x')\, ,
\end{eqnarray*}
where $i,~j~=1,2,3$.

Somehow 3+1 dimensions turns out to be the natural setting for this approach. Till now we have used the Hamiltonian formalism and hence singled out time. If we wish to rework in a fully relativistic covariant 4- dimensional approach one may have to take into account the commutation relations which may arise from 4 manifold invariants. See [16], for example. 

Taking only the Gauss linking number into account would keep time a $c$-number and replace other operators by objects like
$$
\frac{\partial \hat{A}_i(x)}{\partial \hat{x}_j(x)}~=~ \int\int \int [\hat{A}_i(x), \hat{x}^*_j(x')] d^3x' \, .
$$ 
Hence to write the various equations like Maxwell, Dirac etc. at Planck length is straight forward. This should get rid of the problems occurring in defining link invariants by regularizing or thickening. 

We are using some simple ideas from mathematics and physics. Recall that in the special theory of relativity the properties of Maxwell 's equations were used to conclude that the structure of spacetime had to be modified. In this paper we assume fields are fundamental. There are two ways of viewing motion in space. We could take the particle view point and deal with elements like $dx_i$. We could alternatively take the field view point and study the $A_i(x)$. In this paper we begin with the field theory commutator which encodes the field view point and derive the commutation relations for the particle view point. This leads to quantization of volume. 

To the more mathematically minded we are using the philosophy of the Gelfand-Naimark theorem ([14], [15]) which says that the properties of a space are encoded in the algebra of functions residing in the space. The non-commutative nature of the product of the functions leads to noncommutative structure for length and area. 

To carry out the program of quantum gravity is now possible as there is a natural regulator. We can take the approach that quantum field theory is fundamental and the geometrical trappings as incidental [13] or start with the traditional approach of defining metrics and connections taking into account these new commutation relations. A heuristic argument (for 3+1 dimension space) is as follows. We begin with Newton 's law of gravitation 
${\displaystyle \frac{Gm_1m_2}{r^2}}$ 
which can be rewritten using the Planck length $L$ as 
${\displaystyle 
\frac{L^2}{\hbar c} \frac{m_1c^2 m_2 c^2}{r^2}}
$
which goes over in field theory to 
$
{\displaystyle \frac{L^2}{\hbar c} T \Delta T
}$
which is the interaction of the energy momentum tensor mediated by spin-2 gravitons. This is in analogy to electromagnetic theory where it is like 
${\displaystyle \frac{e^2}{\hbar c} J \Delta J}$.
This is the linear approximation. The full theory is got by several methods all of which lead to the Einstein 's theory [13].
The recent advances are essentially going over to the $(A_i,~E_j)$ formalism. It can be seen that scaling length by $L$ leads to scaling the field by ${\displaystyle \sqrt{\frac{L}{\hbar c}}}$. This is an additional evidence that quantization should occur as there is a scale for length occurring naturally. The commutation relations proposed are therefore suitable. 

Alternatively one could rework the program in the fashion Einstein did it.
The entire structure of Riemannian geometry will have to be reworked taking into account the fact that the new commutation relations will affect the mathematics and physics. The notion of connection, metric and all such features will have to be generalized and this would lead to the full theory of physics at the Planck length and quantum gravity. This will be done in a later paper.

\noindent {\bf Acknowledgements:} Most of the ideas used in the paper arose because of course work given when I was doing my Ph.D. at the Institute of Mathematical Sciences, Chennai, India. I take this opportunity to place on record my gratitude to my teachers there. The influence of Professor E.C.G. Sudarshan permeates this paper: he has repeatedly emphasized the importance of Lie groups as the language of physics and always pointed out the formal analogies between classical and quantum equations. Professor G. Rajasekaran asked me to relook at the Bohr-Rosenfeld relations in a course on gravity and also has always emphasized the field theory view point of gravity. I hope this analysis a decade later satisfies him. Professor R. Simon and Professor N.D. Haridass gave useful courses on Lie groups and differential geometry respectively. I have benefitted from frequent discussions with Professor H.S. Sharatchandra, teacher extraordinary.  I would like to thank Mr. K.B. Chandrasekhar and Professor C.N. Krishnan who have supported me in every possible way.

This paper is a belated offering to Bhagawan Sri Sathya Sai Baba on the occasion of his $75^{th}$ birthday: 
\begin{center}
{\it Manasa bhajare Guru charanam\\
Dushtara bhavasagara tharanam}\\ 
\end{center}

\noindent {\bf References}

These references are indicative and not exhaustive.
\begin{description}
\item {\it For the Bohr-Rosenfeld analysis}:-
\item{[1]} Niels Bohr Collected Works, Vol 7, Foundations of Quantum Physics II (1993-58), Editor Jorgen Kalckar, Elsivier.
\item{[2]} Quantum theory of Radiation, W. Heitler, Third Edition, 1960, Claredon Press, Oxford.

\noindent {\it For cocycles, Weyl-Heisenberg groups and physics}:- 
\item{[3]} Lie groups, Lie algebras, Cohomology and some applications in physics, Jose A. de Azcarraga and Jose M. Izquierdo, Cambridge University Press, 1998.
\item{[4]} Chern-Simmons terms and cocycles in physics and mathematics , R. Jackiew in Quantum field theory and Quantum statistics, Volume II, A. Batalin, C.J. Isham and G.A. Vilkovisky Editors, Adam Hilger, 349-378.
\item{[5]} Geometric quantization, N.M.J. Woodhouse, Second Edition, Claredon Press, Oxford, 1994.\\
\noindent {\it For Heisenberg's conjecture}:-
\item{[6]} The self-energy of the electron, W. Heisenberg, Zeitschrift fur Physik, 65: 4-13 (1930) Reprinted in Early Quantum Electrodynamics: A source book, Arthur I. Miller, Cambridge University Press, 1995.

\noindent {\it For Gauss linking number}:-
\item{[7]} Topology and physics- a historical essay, Charles Nash in History of Topology, Editor I.M. James, 1999, Elsivier.
\item{[8]} Gauss linking number and electromagnetic uncertainty principle, A. Ashtekar and A. Corichi, hep-th/9701136.

\noindent {\it For (A, E) formalism and loop quantum gravity}:-
\item{[9]} A new Hamiltonian formulation of general relativity, Phys. Rev. D, 36 (6), 1587-1603, 1987. 
\item{[10]} Quantum Mechanics, L.I. Schiff, Third Edition, Mc Graw-Hill Kogakusha.
\item{[11]} 3+1 dimensional Yang-Mills theory as a local theory of evolution of metrics on 3 manifolds, P. Majumdar and H.S. Sharatchandra, Phys. Let. B, 491 (2000) 199-202. hep-th/0008077.
\item{[12]} Loop Quantum Gravity, C. Rovelli in LIVING REVIEWS (Electronic Journal) gr-qc/9710008.

\noindent{\it  For various approaches to get the Einstein theory starting from field theory}:-
\item{[13]} See the Foreword (Derivation of the Einstein field equation), Feynman Lectures on Gravitation, R. Feynman, F.B. Moringo, W.G. Wagner, Editor B. Hatfield, Perseus Books.

\noindent {\it For Gelfand-Naimark theorem}:-
\item{[14]} Introduction to topology and modern analysis, G.F. Simmons, Mc Graw-Hill, 1963.
\item{[15]} Noncommutative geometry, A. Connes, I.H.E.S./M/93/54.

\noindent {\it For possible generalization to 4 dimensions}:-
\item{[16]} A gauge field approach to 3- and 4- manifold invariants, B. Broda, q-alg/9511010. 
\end{description}
\end{document}